\documentclass{PoS}

\title{Search for the decays $B^+\rightarrow D^+ K^{(*)0}$}

\ShortTitle{Search for the decays $B^+\rightarrow D^+ K^{(*)0}$}

\author{\speaker{Xavier Prudent}\\    
        Inst. fuer Kern- und Teilchenphysik (IKTP)-Technische Universitaet Dresden\\
        E-mail: \email{prudent@physik.tu-dresden.de}}

\abstract{We report a search for the rare decays
$B^+\rightarrow D^+ K^0$ and $B^+\rightarrow D^+ K^{*0}$ in an event sample of
approximately $465$ million $B\overline{B}$ pairs collected with the {\it BaBar}
detector at the PEP-II asymmetric-energy $e^+e^-$ collider at SLAC. We
find no significant evidence for either mode and we set 90\%
probability upper limits on the branching fractions of $BF(B^+\rightarrow D^+ K^0) < 2.9\times 10^{-6}$ and $BF(B^+\rightarrow D^+ K^{*0}) <
3.0\times 10^{-6}$~\cite{cite:denis}.}

\FullConference{35th International Conference of High Energy Physics\\
		 July 22-28, 2010\\
		 Paris, France}

\begin{document}

\section{Introduction}

\begin{figure}
\begin{center}
\includegraphics[width=.4\textwidth]{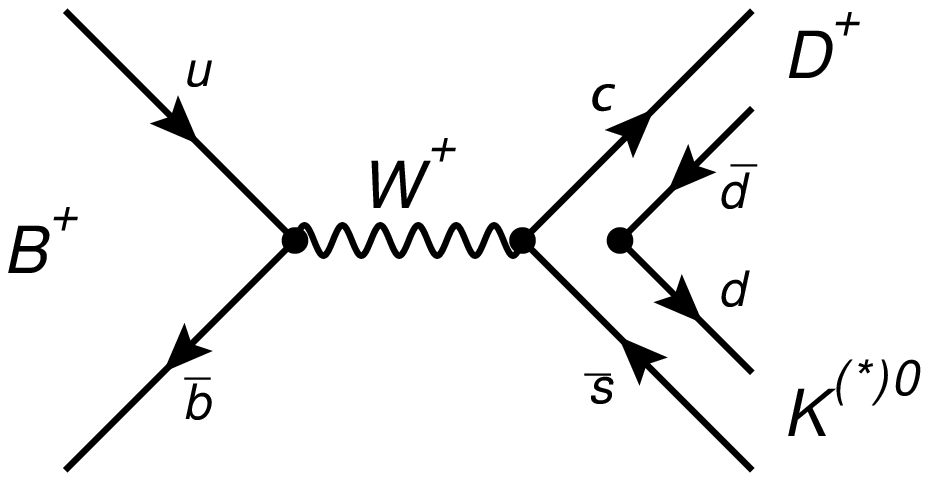}
\includegraphics[width=.4\textwidth]{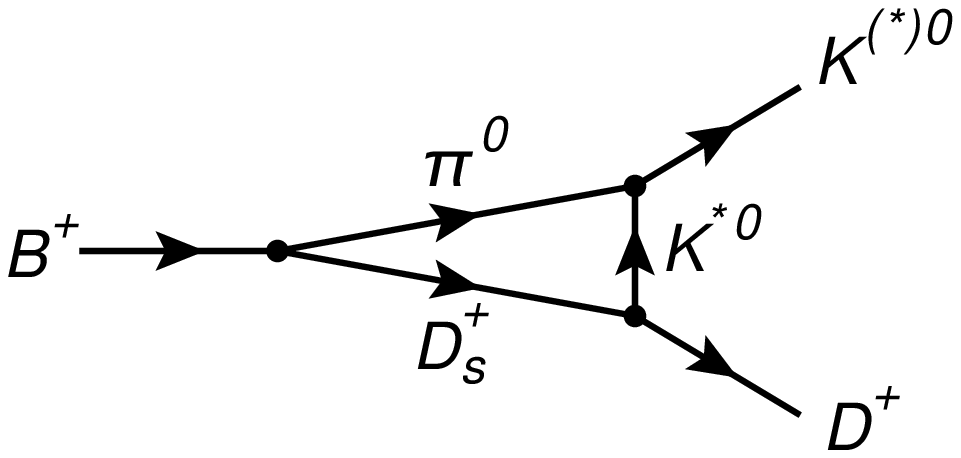}
\caption{\label{diagr}Annihilation diagram for the decay
$B^+{\rightarrow}D^+ K^{(*)0}$ (left) and hadron-level diagram (right)
for the rescattering contribution to $B^+\rightarrow D^{+} K^{(*)0}$ via
$B^+ \rightarrow D_s \pi^0$.}
\end{center}
\end{figure}

Charged $B$ meson decays like $B^+ \rightarrow D^+ K^{(*)0}$ are dominated by weak annihilation diagrams, for which no reliable estimates for the decay rates exist because of soft gluons exchange. In particular annihilation amplitudes cannot be evaluated with the commonly-used
factorization approach~\cite{cite:buras}. Such annihilation amplitudes are suppressed by $\lambda^5$
where $\lambda$ is the sine of the Cabibbo
angle~\cite{ cite:buras,cite:gronau-anni}. So far, no pure annihilation hadronic diagram has been observed, 
and such amplitudes are usually neglected in the measurement of $V_{ub}$. Their branching fractions could be enhanced by so-called
rescattering effects (see Fig.~\ref{diagr}), up to $\lambda^4$~\cite{cite:gronau-anni},
rendering the rate comparable to the isospin-related $B^{0} \rightarrow D^0
K^{(*)0}$ decay rate of approximately $5 \times 10^{-6}$.

None of the modes studied in this note has been observed so far, and a 90\% confidence
level upper limit on the branching fraction ${\ensuremath {{\cal
B}(B^+{\rightarrow}D^+ K^0})} < 5\times10^{-6}$ has been established by
$BaBar$~\cite{cite:polci}.  No study of $B^+\rightarrow D^+K^{*0}$ has
previously been published. The results presented here are obtained with 426~fb$^{-1}$ of data
collected at the $\Upsilon(4S)$ resonance with the {\it BaBar} detector at the
PEP-II asymmetric $e^+e^-$ collider~\cite{cite:pep2} corresponding to
$465 \times 10^6\ B\overline{B}$  pairs ($N_{B\overline{B}}$). An additional 44.4~fb$^{-1}$
of data (``off-resonance'') collected at a center-of-mass (CM)
energy 40~MeV below the $\Upsilon(4S)$ resonance is used to study
backgrounds  from $e^+ e^- \rightarrow q\overline{q}$ ($q=u,\,d,\,s,$ or $c$)
processes, which we refer to as continuum events. The {\it BaBar}  detector is described in detail elsewhere
\cite{cite:det}.

\section{Event Reconstruction and Selection}

The $D^+$ mesons are reconstructed in the modes $D^+ \rightarrow K^-\pi^+\pi^+$ ($K\pi\pi$), $D^+ \rightarrow K_S \pi^+$
($K_S \pi$), $D^+ \rightarrow K^-\pi^+\pi^+ \pi^0$ ($K\pi\pi\pi^0$) and $D^+
\rightarrow K_S \pi^+ \pi^0$ ($K_S \pi\pi^0$) for the decay channels $B^+\rightarrow D^+ K^0$ ($DK$). Only the first two modes are used for the $B^+
\rightarrow D^+K^{*0}$ decay channel ($DK^*$). The event selections are optimized by maximizing $S/\sqrt{S+B}$, where $S$ and $B$ are the expected signal and background yields, using Monte Carlo (MC)
simulations and off-resonance data. The signal branching fraction is taken to be $5\times
10^{-6}$.

The charged kaons are required to satisfy kaon identification
criteria obtained from the combination of information from the Cherenkov light and the tracking detectors.
Kaons and pions must satisfy $p_K>200~$MeV$/c$ and $p_{\pi}>150~$MeV$/c$, where $p$ is the momentum in lab frame.
The invariant mass of the $D^+$ candidates is required to stand within 10 to 22~MeV$/c^2$ (depending on the channel) of the nominal mass~\cite{cite:PDG}. The $K_S$ candidates are reconstructed from $\pi^+\pi^-$ pairs with invariant mass within 5 to 7 MeV$/c^2$ of
the nominal $K_S$ mass~\cite{cite:PDG}. 
We define $\alpha_{K_S}(B^+)$ as the angle between the momentum vector of the
$K_S$ candidate and the vector connecting the $B^+$ and $K_S$ decay
vertices. The prompt $K_S$ candidates from the $B^+\rightarrow D^+K_S $
decay must fulfill $\ln (1-\cos \alpha_{K_S }(B^+))<-8$ and $\ln(1-\cos
\alpha_{K_S }(D^+))<-6$, where $\alpha_{K_S }(D^+)$ is defined in a
similar way. The $\pi^0$ candidates are reconstructed from photon pairs $\gamma\gamma$ with invariant mass $m(\gamma\gamma)$ within 10 to 12~MeV$/c^2$ of the nominal $\pi^0$ mass~\cite{cite:PDG}.
These pairs must satisfy $E(\gamma)>70$~MeV, $E(\gamma\gamma)>200$~MeV, $P_{CM}(\gamma\gamma)>400$~MeV, 
where $E$ and $P_{CM}$ are respectively the energy and the momentum in the CM frame. The $K^{*0}$ candidates are reconstructed in $K^{*0} \rightarrow K^+ \pi^-$ with the invariant mass liying within 40~MeV$/c^2$ of the nominal $K^{*0}$
mass~\cite{cite:PDG}. We define $\theta_{H}$ as the angle
between the direction of flight of the charged $K$ and the direction
of flight of the $B$ in the $K^{*0}$ rest frame, and require $|\cos\theta_{\rm
H}|>0.5$. The $B^+$ candidates are reconstructed by combining one $D^+$ and one
$K_S $ or $K^{*0}$ candidate, constraining them to originate from a
common vertex. We define $\theta_{B}$ as the $B$
polar angle with respect to the beam axis in the CM frame, and require $|\cos\theta_{B}|$ to be smaller than 0.76 to 0.86 depending on the channels. Using the precise knowledge of the $e^+e^-$ beams energies and the energy conservation in the two-body decay $\Upsilon(4S)\rightarrow B\overline{B}$, we define the
beam-energy substituted mass $m_{ES}$ and the energy difference $\Delta E$:
$$
m_{ES}\equiv\sqrt{((E^{*2}_{CM}/c^2)/2-p^{*2}_B},\ \Delta E \equiv E^{*}_B-E^*_{CM}/2,
$$
where $E$ and $p$ are energy and momentum. We retain candidates with $|\Delta E|$ value smaller than 19 to 25~MeV and $m_{ES}$ in the range $[5.20,5.29]~$GeV$/c^2$. 
Multiple $B$ candidates are eliminated with selections on $D^+$ mass or $\Delta E$ distribution. The dominant background comes from continuum events, characterized by a
jet-like topology, which can be described with these variables defined in the CM:
the cosine of the angle between the $B$ thrust axis and the thrust axis of all the other tracks and  energy deposits of the event, where the thrust axis is defined as the direction that maximizes the sum of the longitudinal momenta of all the particles, the event shape moments
$L_0=\sum_{i} p_i$, and $L_2 =\sum_{i} p_i |\cos \theta_i|^2$, where
the index $i$ runs over all tracks and energy deposits in the rest
of the event; $p_i$ is the momentum and $\theta_i$ is the angle to the $B$ thrust axis. We also use $|\Delta t|$, the
absolute value of the time interval between the two
$B$ decays~\cite{Aubert:2002rg}. These four variables are combined in a Fisher discriminant
$F$~\cite{cite:Fish}, whose coefficients are determined with samples of simulated signal and
continuum events, and validated using off-resonance data.
For the $K\pi\pi$ mode, events are classified according to their flavor-tagging
category~\cite{Aubert:2002rg} (lepton, kaon or other) and fitted simultaneously.
The $B\overline{B}$ background is divided into two components according to their distribution in the signal region: non-peaking and peaking. The peaking backgrounds are rejected using the $K_S$ helicity angle $\theta_{K_S }$ with $|cos(\theta_{K_S })|>0.8$ or $0.9$ depending on the channel. 
Based on MC studies, atmost one $B\overline{B}$ peaking
background event per mode is expected in the signal region. The charmless background is evaluated from data using the $D^+$
sidebands and found to be negligible.

\section{Fit Procedure}

\begin{figure}
\begin{center}
\includegraphics[width=.35\textwidth]{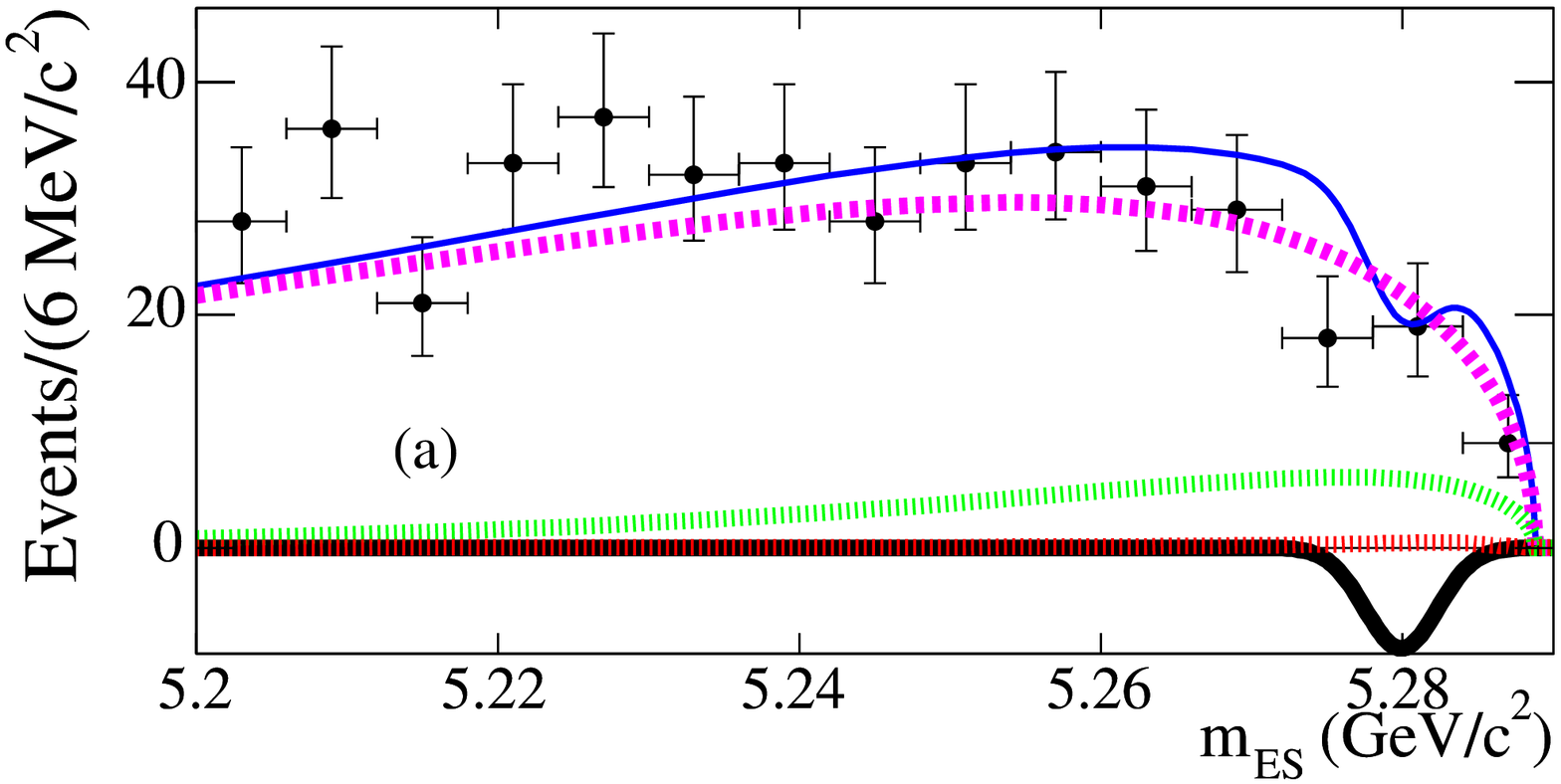}
\includegraphics[width=.35\textwidth]{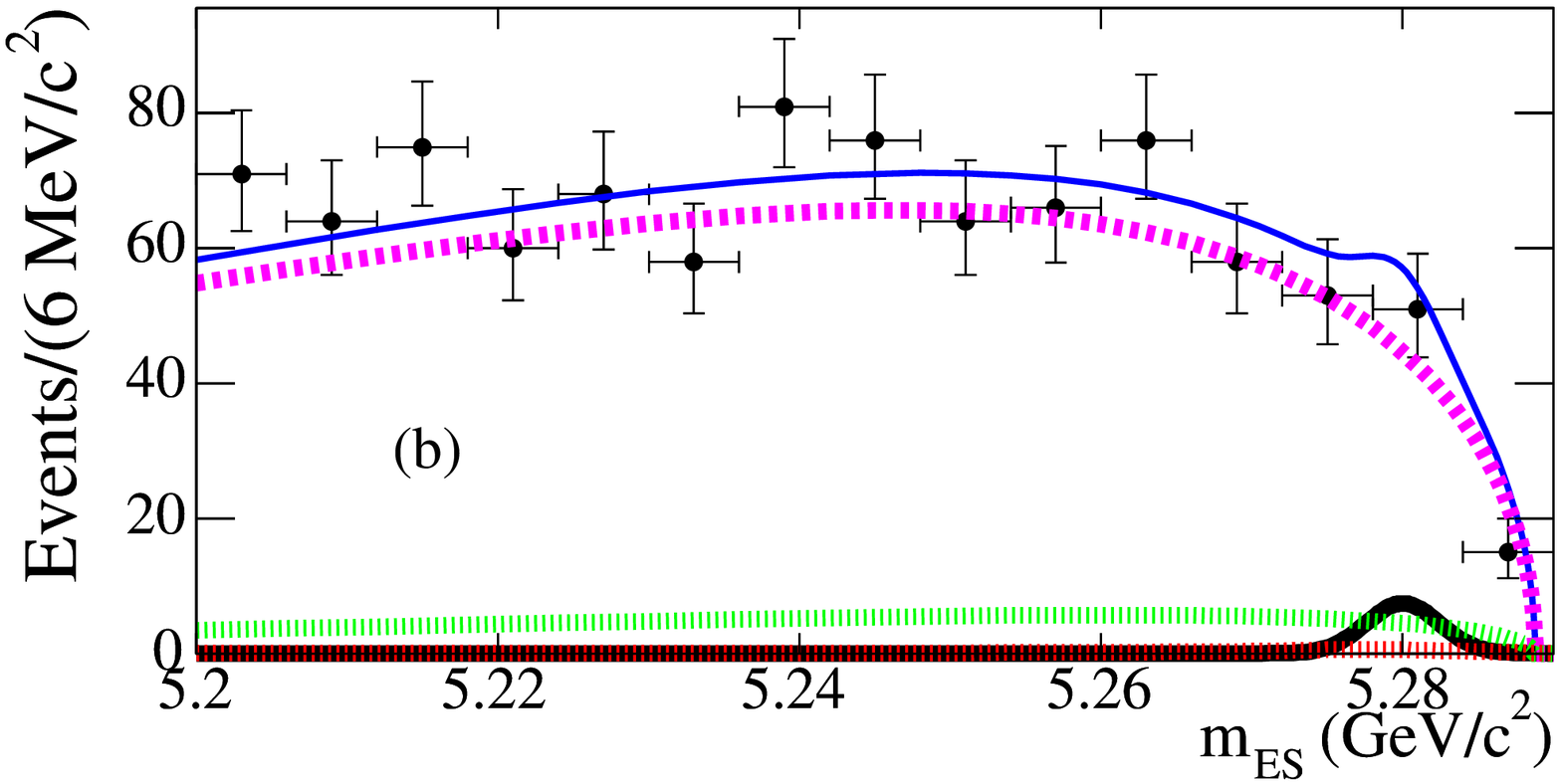}
\includegraphics[width=.35\textwidth]{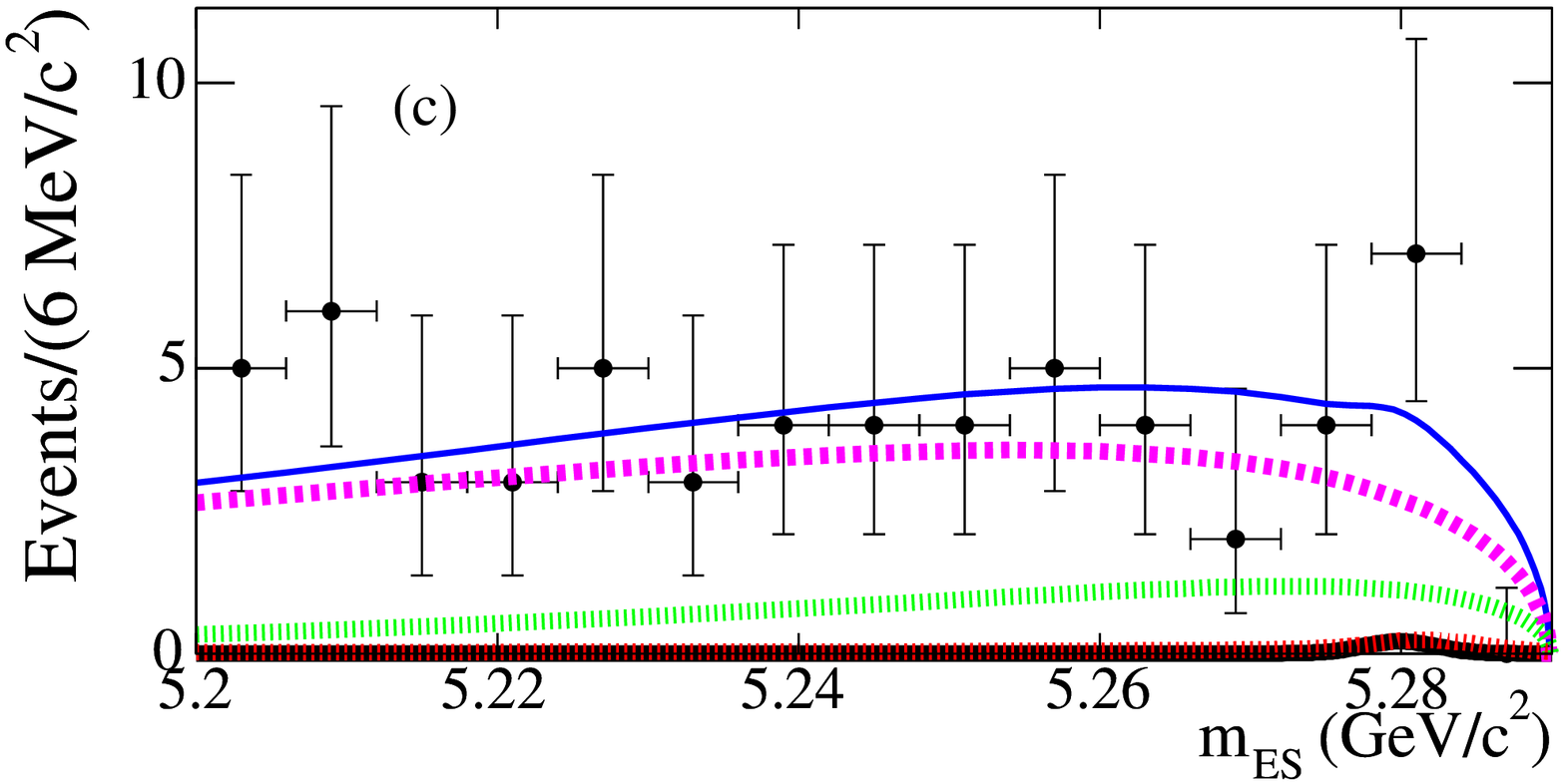}
\includegraphics[width=.35\textwidth]{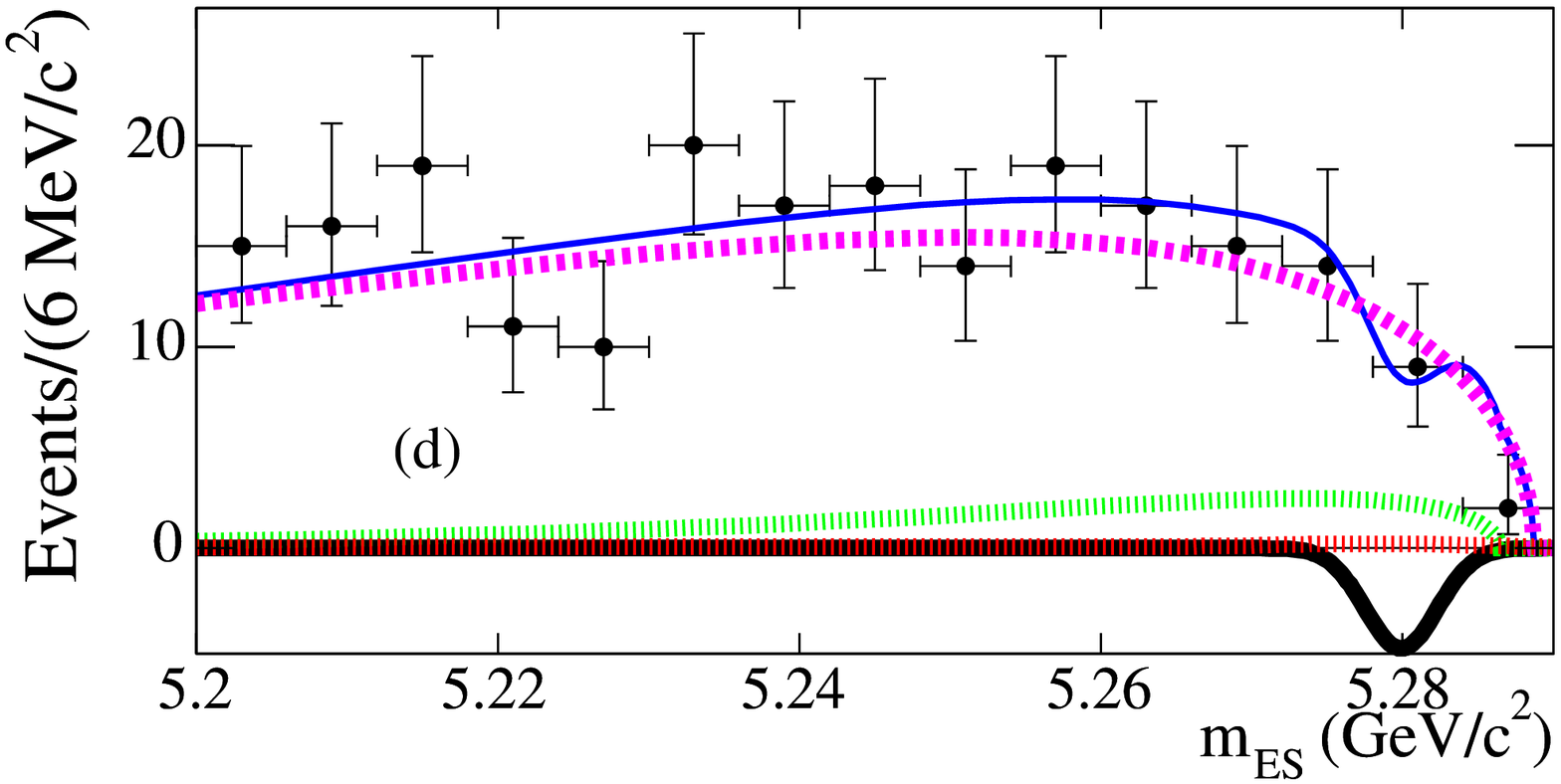}
\includegraphics[width=.35\textwidth]{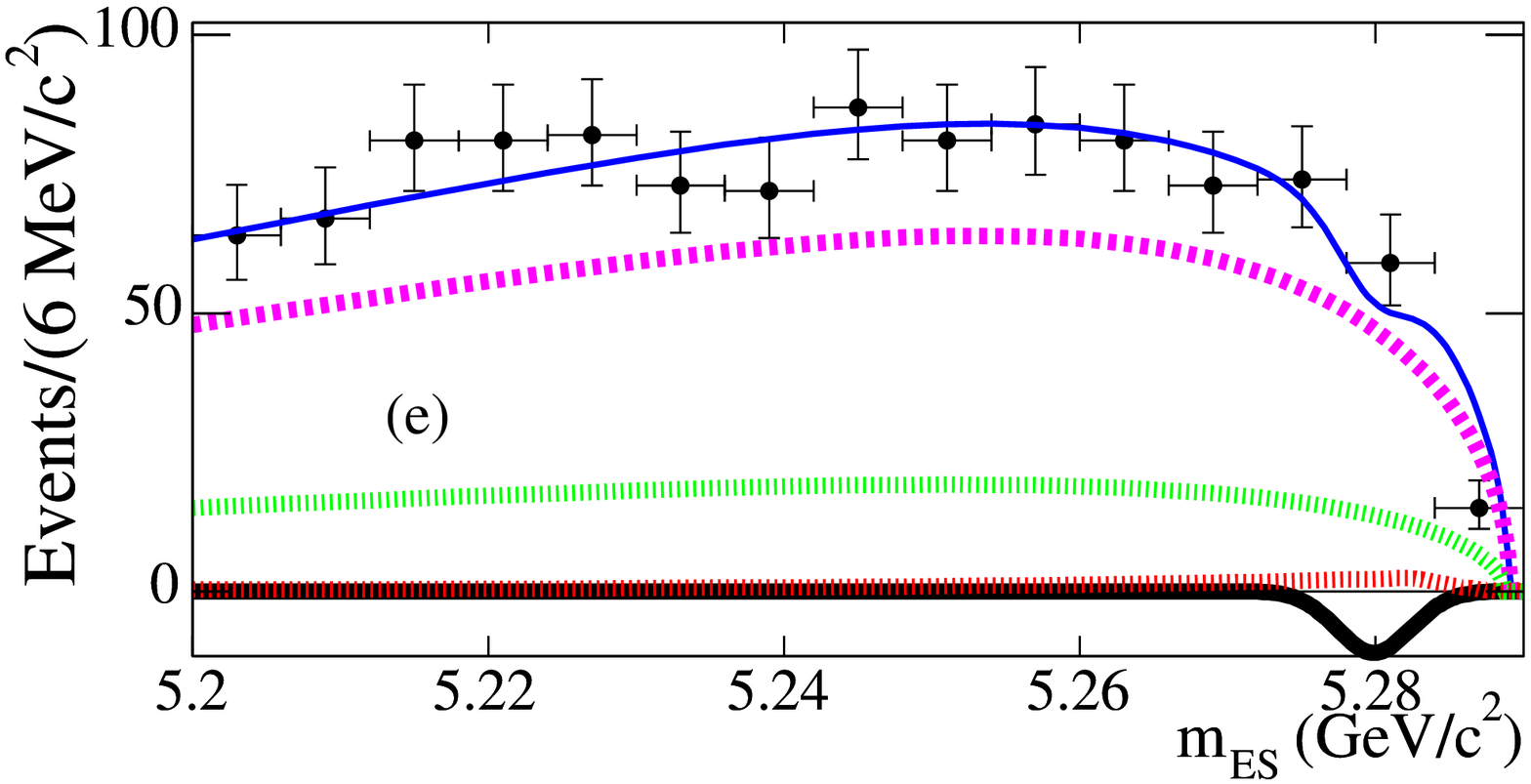}
\includegraphics[width=.35\textwidth]{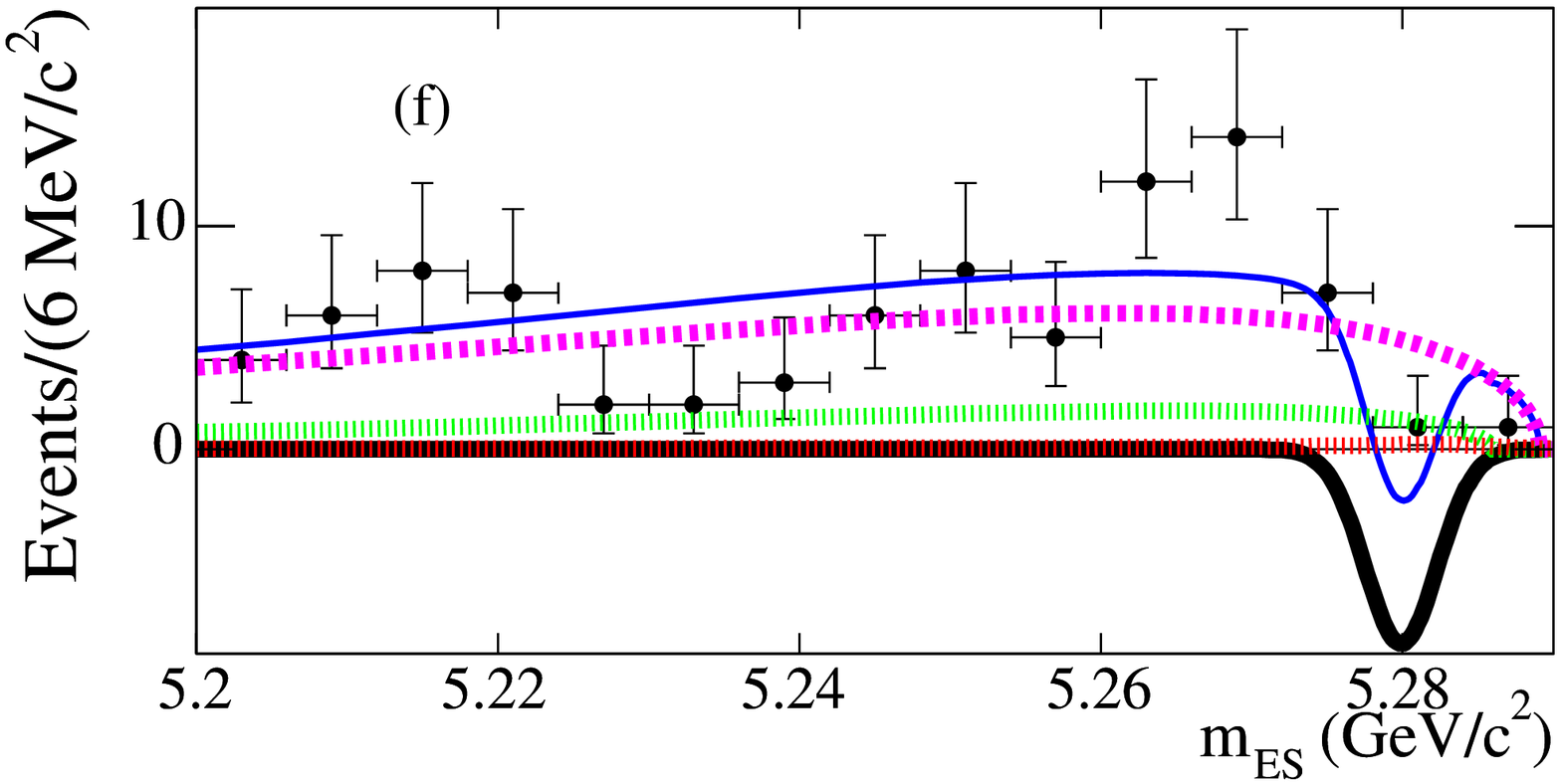}
\caption{From top left to bottom right: $m_{ES}$
projection with $F>0$ for $K \pi \pi$, $K \pi \pi \pi^0$, $K_S 
\pi$, and $K_S  \pi \pi^0$ for $B^+ \rightarrow\ D^+ K_S $ and
$K \pi \pi$ and $K_S  \pi$ for $B^+ \rightarrow\ D^+ K^{*0}$. Data are black dots with error bars, the
different fit components are : signal (black curve),
non-peaking $B\overline{B}$ (green), continuum (magenta) and
$B\overline{B}$\ peaking background (red) and the total pdf (blue).\label{fig:shapes_dkst}}
\end{center}
\end{figure}

The signal and background yields are extracted with an unbinned maximum likelihood fit of $m_{ES}$ and $F$, assuming from simulation studies the correlations between $m_{ES}$ and $F$ to be negligible.
For $m_{ES}$ the signal is modeled with a Gaussian function, the continuum and non-peaking $B\overline{B}$ background are described by two ARGUS
functions~\cite{cite:argus}: $A(x) = x\sqrt{1-(x/x_0)^2}\cdot
\exp(c(1-(x/x_0)^2))$, where $x_0$ is the maximum value of $x$ and $c$ accounts for the shape of the distribution and are determined from data for the continuum. All other PDF parameters are derived from
the simulated events. The peaking $B\overline{B}$ background is modeled with a Crystal
Ball function~\cite{cite:CB} which is a Gaussian
modified to include a power-law tail. The peaking background yield is fixed from the PDG branching fractions~\cite{cite:PDG}. The signal yield determined by the fit ($N_{sig}$) is
used to calculate the branching fraction (BF): $BF =N_{sig}/(N_{B^+}\cdot\epsilon_{sig}\cdot BF_{sec})$, where $N_{B^+}$ is the total number of charged $B$ mesons
in the data sample, $BF_{sec}$ is the BF is of the secondary decay channels of the $D$
and $K_S$, and $\epsilon_{sig}$ is the
signal reconstruction efficiency measured in MC. The fit procedure is validated using toy MC studies and no biases
of the fit model were found. The fit model was tested using full MC sample with and without signal events.
The results of the fit to the data are reported in
Table~\ref{tab:fit} for each $D$ channel. The background
yields are close to the expectations and the errors obtained on the
branching fractions are in good agreement with the values found with the toy study. The leading contribution is
obtained from the $K\pi\pi$ mode. The Fig.~\ref{fig:shapes_dkst} gives the fit projection for $m_{ES}$, after
requiring $F>0$, to visually enhance any possible signal.

\begin{table*}[htb]
\footnotesize
\renewcommand{\arraystretch}{1.2}
\begin{center}
\caption{Branching fraction (BF) measured in units of $10^{-6}$ with
statistical und systematic uncertainties for each channel. $N_{i}$ are the yields of the fitted
species.\label{tab:fit}}
\begin{tabular}{|l|c|c|c|c|}
\hline\hline
 Decay mode & $N_{sig}$ &  $N_{B\overline{B}}$ &  $N_{cont}$  & BF \\
\hline \multicolumn{5}{|l|}{$B^+ \rightarrow D^+ K^0$}\\\hline

~~${K\pi\pi}$ & $-11.9^{+6.7}_{-5.6}$ &$70\pm27$ &$2690\pm57$  & $-4.2^{+2.4}_{-2.0}$(stat.)$^{+1.1}_{-1.3}$(syst.) \\

~~${K\pi\pi\pi^0}$ & $10^{+10}_{-9}$ &$111\pm51$ &$6516\pm94$  & $20^{+20}_{-17}$(stat.)$^{+11.3}_{-11.8}$(syst.) \\

~~${K_S \pi}$ & $0.6^{+5.3}_{-4.5}$ &$20\pm14$ &$381\pm23$  & $0.7^{+15}_{-13}$(stat.)$^{+8.2}_{-9.3}$(syst.) \\

~~${K_S \pi\pi^0}$ & $-6.7^{+4.5}_{-2.8}$ &$36\pm22$ &$1270\pm41$  & $-14^{+9.2}_{-6.2}$(stat.)$^{+9.0}_{-12.5}$(syst.)\\

\hline \multicolumn{5}{|l|}{$B^+ \rightarrow D^+ K^{*0}$ }\\\hline

~~${K\pi\pi}$ & $-15.6^{+8.7}_{-7.1}$ &$463\pm63$ &$6338\pm98$  &$-5.0^{+2.9}_{-2.1}$(stat.)$^{+1.5}_{-1.8}$(syst.) \\

~~${K_S \pi}$ & $-11.4^{+3.5}_{-2.4}$ &$35\pm15$ &$547\pm27$  &$-33^{+10.2}_{-7.0}$(stat.)$^{+6.4}_{-7.4}$(syst.)\\

\hline\hline
\end{tabular}
\end{center}
\end{table*}

\section{Systematic uncertainties}

The uncertainties on the PDF
parameterizations is evaluated by repeating the fit varying the
MC-obtained PDF parameters within their statistical errors, taking
into account correlations among the parameters.
Differences
between the data and MC for the signal PDF shapes are investigated using data
control samples $B^{0}\rightarrow D^+ \pi^-$ and $B^{0}\rightarrow D^+ \rho^-$.
The uncertainty on the continuum background shape is estimated using off-resonance data instead of continuum MC.
The uncertainty on the PDF of the non-peaking $B\overline{B}$\ background is measured by
leaving its parameters free in the fit and taking the difference from the nominal fits as uncertainty. We also considered the uncertainty on signal efficiency due to limited MC statistics. 
Uncertainties on MC-data differences in tracking efficiency,
$K_S$  and $\pi^0$ reconstruction and charged-kaon identification, are estimated by
comparing  data and simulation performance in control samples. The uncertainty on peaking background are estimated by repeating the fit varying the event yields within their statistical errors. 
The uncertainties on the
branching fractions of the sub-decay modes are also taken into
account. The uncertainty on $N_{B\overline{B}}$ has a negligible
effect on the total error. The uncertainties are included by convolving
the individual fit likelihoods with Gaussians of width equal to the systematic uncertainty. The total systematic uncertainties on the BF are given in the Table~\ref{tab:fit} for each channel.

\section{Results for Branching Fractions}

The individual likelihoods for each mode are finally combined to give the average BF's, which are compatible with zero.
We then quote an upper limits at 90\% probability using a Bayesian approach with a flat prior for the BF:
\begin{eqnarray*}
BF(B^+ \rightarrow D^+ K^0) < 2.9\times 10^{-6},\ 
BF(B^+ \rightarrow D^+ K^{*0}) < 3.0\times 10^{-6}.
\end{eqnarray*}


\begin{thebibliography}{99}
\bibitem{cite:denis} B. Aubert {\it et al.} ($BaBar$ Collaboration), Phys. Rev. D { \bf 82}, 092006 (2010).
\bibitem{cite:buras} A.~J.~Buras and L.~Silvestrini, Nucl. Phys. B {\bf 569}, 3 (2000).
\bibitem{cite:gronau-anni} B. Blok, M. Gronau, and J.L. Rosner, Phys. Rev. Lett. {\bf 78} 3999 (1997).
\bibitem{cite:polci} B. Aubert {\it et al.} ($BaBar$ Collaboration), Phys. Rev. D { \bf 72}, 011102 (2005).
\bibitem{Aubert:2002rg}  B.~Aubert {\it et al.}  ($BaBar$ Collaboration),  Phys.\ Rev.\  D {\bf 66}, 032003
(2002).
\bibitem{cite:pep2} {\it PEP II - An Asymmetric B Factory, Conceptual Design Report}, SLAC-418, LBL-5379 (1993).
\bibitem{cite:det} B. Aubert {\it et al.} ($BaBar$ Collaboration), Nucl. Instr. and Methods A {\bf 479}, 1 (2002).
\bibitem{cite:PDG} C. Amsler {\it et al.} (Particle Data Group), Phys. Lett. B {\bf 667}, 1 (2008).
\bibitem{cite:Fish} R.~A.~Fisher, Annals Eugen. {\bf 7}, 179 (1936).
\bibitem{cite:argus} H.~Albrecht {\it et al.} (ARGUS Collaboration), Z. Phys. C {\bf 48}, 543 (1990).
\bibitem{cite:CB} J.~E.~Gaiser, Ph.D. thesis, Stanford University [SLAC-R-255] (1982).
\end{thebibliography}
\end{document}